\documentclass[fleqn,usenatbib]{mnras}

\usepackage{newtxtext,newtxmath}
\usepackage[T1]{fontenc}

\DeclareRobustCommand{\VAN}[3]{#2}
\let\VANthebibliography\thebibliography
\def\thebibliography{\DeclareRobustCommand{\VAN}[3]{##3}\VANthebibliography}

\usepackage{graphicx}	% Including figure files
\usepackage{amsmath}	% Advanced maths commands

\title[Brown Hamiltonian]{Extensions of Brown Hamiltonian--III. Applications to irregular satellites of giant planets}

\author[Lei et al.]{
Hanlun Lei,$^{1,2}$\thanks{E-mail: leihl@nju.edu.cn}
Xiaoyan Leng,$^{1,2}$
Evgeni Grishin$^{3,4}$\\
$^{1}$School of Astronomy and Space Science, Nanjing University, Nanjing 210023, China\\
$^{2}$Key Laboratory of Modern Astronomy and Astrophysics in Ministry of Education, Nanjing University, Nanjing 210023, China\\
$^{3}$School of Physics and Astronomy, Monash University, Clayton, VIC 3800, Australia\\
$^{4}$OzGrav: Australian Research Council Centre of Excellence for Gravitational Wave Discovery, Clayton, VIC 3800, Australia
}

\date{Accepted XXX. Received YYY; in original form ZZZ}

\pubyear{2026}

\begin{document}
\label{firstpage}
\pagerange{\pageref{firstpage}--\pageref{lastpage}}
\maketitle

\begin{abstract}
Irregular satellites, orbiting at large distances from their host planets, are subject to strong solar perturbations that render long-term orbital predictions particularly challenging. Building upon the extended Brown Hamiltonian framework developed in Paper I, we introduce the modified Lidov integral ($C_{\rm ZLK}$) as a practical diagnostic index to characterize the dynamical modes. We demonstrate that a satellite is trapped inside the von Zeipel--Lidov--Kozai (ZLK) resonance when $C_{\rm ZLK} < 0$. Applying this criterion to the known population of irregular satellites, we identify 27 candidates in libration. Direct $N$-body simulations confirm 26 of these predictions, with the sole exception of S/2019 S1, whose discrepancy is attributed to its proximity to the separatrix. These results establish $C_{\rm ZLK}$ as a decisive parameter for identifying the ZLK resonance, providing an efficient tool for analyzing the secular dynamics in weakly hierarchical three-body systems.
\end{abstract}


%%%%%%%%%%%%%%%%%%%%%%%%%%%%%%%%%%%%%%%%%%%%%%%%%%

%%%%%%%%%%%%%%%%% BODY OF PAPER %%%%%%%%%%%%%%%%%%

\section{Introduction}
\label{Sect1}

It is generally believed that the irregular satellites of giant planets were primordially captured from heliocentric orbits in the early history of the solar system \citep{saha1993orbits,gladman2001discovery,astakhov2003chaos,cuk2006irregular,nesvorny2007capture,jewitt2007irregular,vokrouhlicky2008irregular,philpott2010three,gaspar2011irregular,gaspar2013irregular,nesvorny2014capture}. Thus, their current dynamical properties offer clues to capture mechanisms and shed light on the evolutionary history of the solar system \citep{peale1999origin,carruba2002inclination,nesvorny2003orbital}. In particular, these objects are characterized by their distant, highly inclined, and often eccentric and/or retrograde orbits. Because of their distant and elongated orbits, they are heavily influenced by solar gravitational perturbations, making their reliable long-term orbital predictions extremely challenging \citep{cuk2004secular,grishin2024irregularI}.

Due to multi-scale architectures of these systems, the hierarchical three-body problem is usually employed. Under the secular (phase-averaged) approximation, various dynamical frameworks have been developed, ranging from the classical secular Hamiltonian model \citep{von1910application,kozai1962secular,lidov1962evolution,naoz2013secular,lithwick2011eccentric} to the Brown Hamiltonian model \citep{brown1936stellarIII,cuk2004secular,breiter2015secular,luo2016double, grishin2018quasi, lei2018modified,tremaine2023hamiltonian} and its recent extension, namely the extended Brown Hamiltonian model \citep{lei2025ExtensionsI}. The applicability of these models depends on the degree of timescale hierarchy, which is measured by the single-averaging parameter $\varepsilon_{\rm SA}$ \citep{luo2016double}, or equivalently, the orbital separation normalized by the Hill radius $\alpha_{\rm H}$ \citep{grishin2017generalized}. Specifically, the secular Hamiltonian model is appropriate for highly hierarchical systems, whereas the classical and extended Brown models are tailored for mild- and weak-hierarchy configurations, respectively \citep{lei2025ExtensionsII}.

In the secular Hamiltonian model, if the inclination exceeds a critical threshold, the third-body perturbation can trigger large-amplitude oscillations in eccentricity and inclination. This dynamical phenomenon is known as the von Zeipel–Lidov–Kozai (ZLK)\footnote{The von Zeipel--Lidov--Kozai (ZLK) mechanism is also widely known as the Kozai--Lidov (KL) or Lidov--Kozai (LK) mechanism, or simply the Kozai mechanism \citep{ito2019lidov}.} effect \citep{von1910application,lidov1962evolution,kozai1962secular,naoz2016eccentric,ito2019lidov}. The ZLK resonance, characterized by the libration of the argument of pericenter, often provides a dynamical configuration conducive to long-term stability. In our solar system, more than ten irregular satellites have been reported in this librating state (see \citealp{grishin2024irregularII} for details). Focusing primarily on S/2000 S5 (Kiviuq), \citet{carruba2004chaos} analyzed the chaotic phase-space dynamics of irregular satellites near the separatrix of the ZLK resonance. They pointed out that chaotic behavior may be driven by resonance overlap between the ZLK resonance and secondary resonances arising from commensurabilities between the Kozai precession/libration frequency and other dynamical frequencies (e.g., the Great Inequality and secular frequencies).

For irregular satellites, the normalized orbital separation $\alpha_{\rm H}$ spans a wide range, reaching values as high as 0.5. Within this regime, the system hierarchy weakens, and the nonlinear effects arising from short-period terms become increasingly pronounced, rendering conventional linear secular models inadequate, especially for highly distant satellites \citep{grishin2024irregularII}. To overcome this limitation, we formulated an extended Brown Hamiltonian in Paper I \citep{lei2025ExtensionsI}, providing a unified and highly accurate framework to capture the complex long-term dynamics. Building upon this theoretical framework, the present work establishes an analytical criterion for identifying the ZLK resonance. Applying this criterion to the known population of irregular satellites, we identify 27 resonance candidates, 26 of which are confirmed through direct $N$-body simulations.

This paper is organized as follows. Following a brief introduction to the extended Brown Hamiltonian framework in Section \ref{Sect2}, Section \ref{Sect3} derives the analytical criterion for the ZLK resonance. Section \ref{Sect4} presents a detailed application to known irregular satellites. Finally, we summarize our main conclusions in Section \ref{Sect5}.

\section{Hamiltonian model}
\label{Sect2}

In the planet--satellite--Sun configuration, the satellite moves around its host planet under the gravitational perturbation of the Sun. Under the invariant plane reference frame, we have developed a high-accuracy Hamiltonian framework in Paper I by incorporating the nonlinear effects of the quadrupole-order third-body disturbing function. In particular, the extended Brown Hamiltonian is expressed in a closed form in terms of the eccentricities of both the inner and outer binaries as follows \citep{lei2025ExtensionsI}:
\begin{equation}\label{Eq1}
{\cal F} = {{\cal F}_{20}} + {\varepsilon _{{\rm{21}}}}{{\cal F}_{21}} + {\varepsilon _{{\rm{22}}}}{{\cal F}_{22}},
\end{equation}
where ${\cal F}_{20}$ is the classical ZLK term \citep{kozai1962secular,lidov1962evolution},
\begin{equation}\label{Eq2}
{{\cal F}_{20}} =  2{e^2}- 5{e_z^2}  + {j_z^2} - \frac{1}{3},
\end{equation}
${\cal F}_{21}$ is the classical Brown Hamiltonian \citep{brown1936stellarIII,soderhjelm1975three,breiter2015secular,luo2016double,tremaine2023dynamics,tremaine2023hamiltonian}, 
\begin{equation}\label{Eq3}
{{\cal F}_{21}} = \frac{3}{8}{j_z}\left( {1 - j_z^2 + 24{e^2} - 15e_z^2} \right),
\end{equation}
and ${\cal F}_{22}$ is the extended Brown Hamiltonian \citep{lei2025ExtensionsI},
\begin{equation}\label{Eq4}
\begin{aligned}
{{\cal F}_{22}} =& \frac{1}{{64}}\left\{{ 8e^2\left(13{e^2} + {22e_z^2 + 4j_z^2 + 120}\right) - 94j_z^2} \right.\\
&\left. { - 3\left[ {95e_z^4 + 6e_z^2\left( {15j_z^2 + 31} \right) + 7j_z^4} \right]} \right\},
\end{aligned}
\end{equation}
with $j_z = \sqrt {1 - {e^2}}{\cos i}$ and $e_z= e \sin i\sin \omega$. Specifically, ${\cal F}_{21}$ and ${\cal F}_{22}$ represent the non-linear effects arising from the short-period terms associated with the outer and inner orbits, respectively. The coefficients ${\varepsilon _{21}}$ and ${\varepsilon _{22}}$ are introduced to measure the significance of the classical and extended Brown corrections, given by 
\begin{equation}
{\varepsilon _{21}} = \frac{1}{{\sqrt 3 }}\alpha _{\rm H}^{3/2}f_{21}(e_\odot),\ \quad
{\varepsilon _{22}} = \frac{1}{3}\alpha _{\rm H}^3f_{22}(e_\odot)
\end{equation}
where $\alpha_{\rm H} = a/r_{\rm H}$ is the semimajor axis ratio to the Hill radius of the host planet $r_{\rm H}=a_{\odot} (m_{\rm p}/3m_\odot)^{1/3}$, with $a_{\odot}$ as the semimajor axis and $e_{\odot}$ as the eccentricity of the outer orbit.  The functions $f_{21}, f_{22}$ are polynomials of $e_\odot$ given in \citet{lei2025ExtensionsII}. For small eccentricity ($e_\odot \ll 1$, valid for all giant planets), they are of order unity and we get
\begin{equation*}
{\varepsilon _{21}} = {\varepsilon _{\rm SA}} \sim \frac{1}{{\sqrt 3 }}{\alpha_{\rm H}^{3/2}} \sim \frac{{{P_{\rm out}}}}{{{t_{\rm ZLK}}}},\quad {\varepsilon _{22}} = {\varepsilon _{\rm SA}^2} \sim \frac{1}{3}{\alpha_{\rm H}^3} \sim \frac{{{P_{\rm in}}}}{{{t_{\rm ZLK}}}},
\end{equation*}
where ${\varepsilon _{\rm SA}}$ is the single-averaging parameter \citep{luo2016double}, $P_{\rm in}$ and $P_{\rm out}$ are the orbital periods of the inner and outer binaries, and $t_{\rm ZLK}$ is the timescale of the classical ZLK oscillation \citep{grishin2017generalized}. We can see that the normalized separation $\alpha_{\rm H}$ is a good parameter to characterize the level of timescale hierarchy. As $\alpha_{\rm H}$ increases, the hierarchical level of system decreases.

\begin{figure}
\centering
\includegraphics[width=1.0\columnwidth]{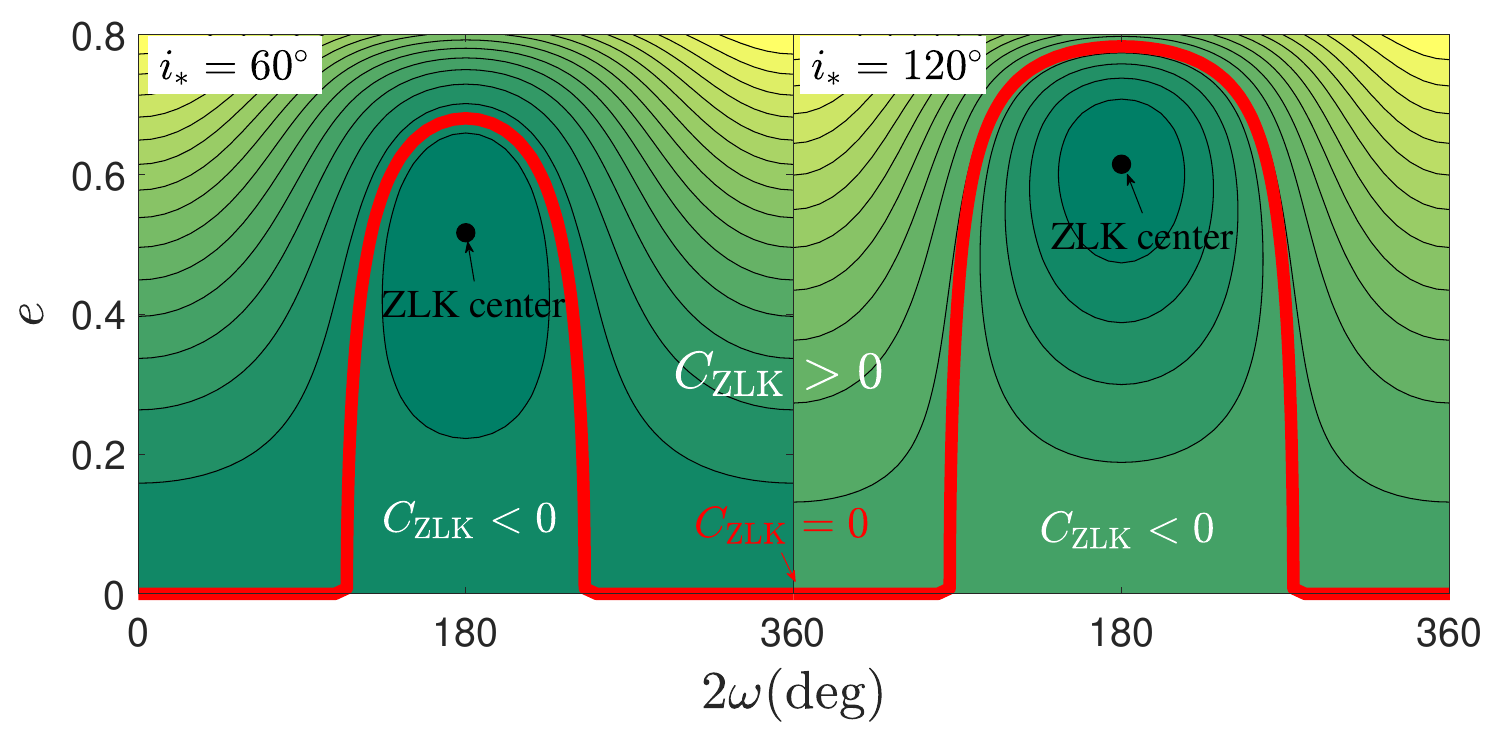}
\caption{Phase portraits (i.e., level curves of the Lidov integral $C_{\rm ZLK}$) under the extended Brown Hamiltonian model with $\alpha_{\rm H}=0.4$ for the cases of $i_* = 60^{\circ}$ (\textit{left panel}) and $i_* = 120^{\circ}$ (\textit{right panel}). The location of the modified ZLK center is marked in black dot. The curve of $C_{\rm ZLK} = 0$ corresponds to the dynamical separatrix between librating regime ($C_{\rm ZLK} < 0$) and circulating regime ($C_{\rm ZLK} > 0$) in the phase space.}
\label{Fig1}
\end{figure}

The Hamiltonian (\ref{Eq1}) determines an integral model, in which the vertical component of angular momentum $j_z$ is a motion integral, characterized by
\begin{equation}\label{Eq5}
j_z = \sqrt{1-e^2}\cos{i} = \cos{i_*},
\end{equation}
where $i_*$ is the mutual Kozai inclination, corresponding to the inclination evaluated at zero eccentricity. In particular, $j_z>0$ stands for the prograde motion and $j_z<0$ represents the retrograde motion.

Under the framework of the extended Brown Hamiltonian, we introduce the modified Lidov integral as follows:
\begin{equation}\label{Eq8}
{C_{\rm ZLK}} = C_{\rm ZLK}^{20} + {\varepsilon _{21}}C_{\rm ZLK}^{21} + {\varepsilon _{22}}C_{\rm ZLK}^{22},
\end{equation}
where
\begin{equation}\label{Eq9}
C_{\rm ZLK}^{20} = {e^2} - \frac{5}{2}e_z^2,\quad C_{\rm ZLK}^{21} = \frac{9}{2}{j_z}\left( {{e^2} - \frac{5}{8}e_z^2} \right),
\end{equation}
and
\begin{equation}\label{Eq10}
\begin{aligned}
C_{\rm ZLK}^{22} =& \frac{1}{{64}}\left\{ {8{e^2}\left( {\frac{{13}}{2}{e^2} + 11e_z^2 + 2j_z^2 + 60} \right)} \right.\\
&\left. { - \frac{3}{2} e_z^2\left[ {95e_z^2 + 6\left( {15j_z^2 + 31} \right)} \right]} \right\}.
\end{aligned}
\end{equation}
The modified Lidov integral ${C_{\rm ZLK}}$ derived in this study accounts for the contributions from the Brown correction (${\cal F}_{21}$) and the extended Brown correction (${\cal F}_{22}$). Thus, it has a wider range of application scenarios. In particular, ${C_{\rm ZLK}}$ reduces to the classical Lidov integral $C^{20}_{\rm ZLK}$ in the limit of $\alpha_{\rm H} \to 0$ \citep{lidov1962evolution,shevchenko2016lidov}.

With the introduction of $C_{\rm ZLK}$, the extended Brown Hamiltonian can be expressed as
\begin{equation}\label{Eq11}
{\cal F} = 2{C_{\rm ZLK}} + \left( {j_z^2 - \frac{1}{3}} \right) + \frac{3}{8}{\varepsilon _{21}}{j_z}\left( {1 - j_z^2} \right) - \frac{1}{{64}}{\varepsilon _{22}}j_z^2\left( {94 + 21j_z^2} \right).
\end{equation}
Similar to the Hamiltonian ${\cal F}$ and $j_z$, the modified Lidov integral ${C_{\rm ZLK}}$ is also a constant of motion. Consequently, $C_{\rm ZLK}$ provides a robust framework for characterizing the ZLK properties. In practice, employing $C_{\rm ZLK}$ is often more advantageous and intuitive than working directly with the Hamiltonian ${\cal F}$ \citep{antognini2015timescales}.

Taking the Jupiter–satellite–Sun system as a representative case, Figure \ref{Fig1} displays the level curves of $C_{\rm ZLK}$ (i.e., phase portraits) in the $(2\omega, e)$ plane for $i_* = 60^{\circ}$ and $i_* = 120^{\circ}$. The (pseudo) phase space is clearly partitioned into libration and circulation regimes by the dynamical separatrix, defined by $C_{\rm ZLK} = 0$.

\section{Analytical criterion}
\label{Sect3}

As illustrated in Figure \ref{Fig1}, the dynamical separatrix is specified by $C_{\rm ZLK} = 0$, which corresponds to the level curve passing through the saddle point at $e=0$. Consequently, $C_{\rm ZLK}$ can be used to separate the libration and circulation regimes: (a) if $C_{\rm ZLK} < 0$, the argument of pericenter $\omega$ librates around $\pi/2$ or $3\pi/2$; (b) if $C_{\rm ZLK} > 0$, $\omega$ undergoes full circulation over the range $[0, 2\pi]$. This behavior remains consistent with the classical ZLK theory \citep{lidov1962evolution, broucke2003long, antognini2015timescales, shevchenko2016lidov}. Within the framework of the classical ZLK model (${\cal F}_{20}$), the partitioning of the parameter space $(c_1 = j_z^2, c_2 = C_{\rm ZLK}^{20})$ into libration and circulation regimes is represented by the Lidov diagram \citep{lidov1962evolution,broucke2003long,shevchenko2016lidov}. This representation is helpful in understanding the phase-space topology and the associated ZLK oscillations.

\begin{figure}
\centering
\includegraphics[width=1.0\columnwidth]{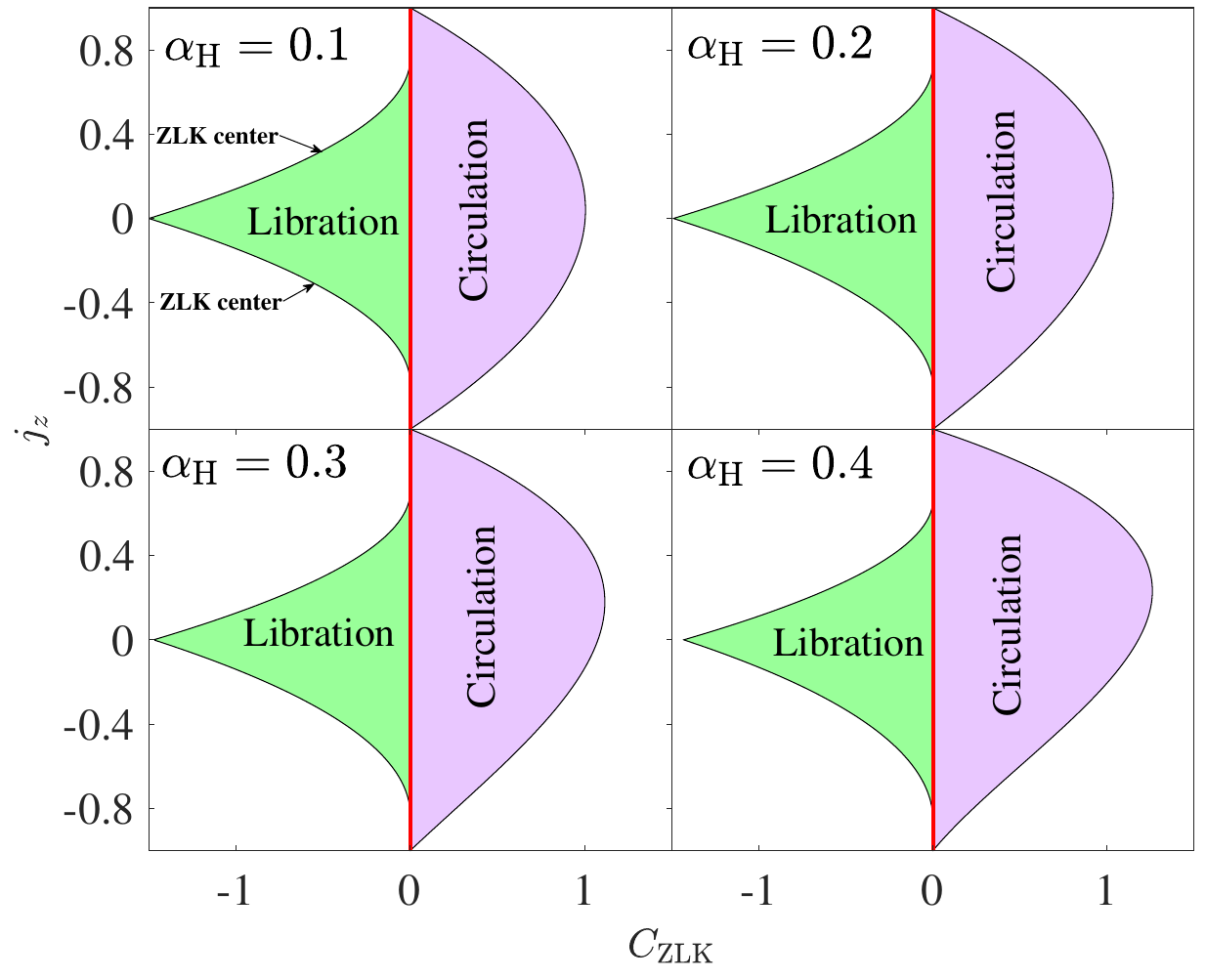}
\caption{Distribution of librating and circulating regimes in the parameter space of $(C_{\rm ZLK},j_z)$, under the extended Brown Hamiltonian model with different levels of $\alpha_{\rm H}$. The separatrix characterized by $C_{\rm ZLK}= 0$ is marked in red lines. The unfilled regions represent dynamically forbidden zones.}
\label{Fig2}
\end{figure}
 
Under the extended Brown Hamiltonian framework, we produce the associated Lidov diagram in the parameter space of $(C_{\rm ZLK}, j_z)$, as shown in Figure \ref{Fig2}. The libration and circulation domains remain partitioned by the $C_{\rm ZLK} = 0$ boundary (see red lines). As the separation parameter $\alpha_{\rm H}$ increases, the asymmetry with respect to the $j_z = 0$ (polar orbit) axis becomes progressively more pronounced due to the classical Brown correction term, ${\cal F}_{21}$, which is an odd function of $j_z$ \citep{lei2025ExtensionsII}.

For a given $j_z$, the Lidov integral $C_{\rm ZLK}$ reaches its minimum at the ZLK resonance center. Consequently, these centers are localized along the left-hand boundaries of the libration regimes, as illustrated in Figure \ref{Fig2}. The maximum eccentricity is evaluated along the dynamical separatrix where $2\omega = \pi$, implying that the points of peak eccentricity lie on the curve of $C_{\rm ZLK} = 0$ (see the red lines).

As illustrated in Figure \ref{Fig2}, the condition for ZLK resonance within the extended Brown Hamiltonian framework can be stated as follows: a satellite is trapped inside the ZLK resonance only if it satisfies $C_{\rm ZLK} < 0$. This criterion provides an analytical tool for rapid identification of ZLK librating candidates for satellites within the Solar system.

It is important to emphasize that this analytical criterion may become invalid in the presence of additional perturbations (e.g., from other planets), since the $C_{\rm ZLK}$ integral is strictly derived for restricted hierarchical three-body systems. Moreover, the criterion is prone to failure for satellites situated near the ZLK separatrix, where chaotic dynamics dominate \citep{carruba2004chaos}.

\section{Applications to irregular satellites}
\label{Sect4}

In this section, we apply the analytical criterion for classifying libration and circulation to irregular satellites of giant planets in the Solar System. As of October 11, 2025, a total of 413 satellites have been identified orbiting the giant planets, comprising 97 of Jupiter, 272 of Saturn, 28 of Uranus, and 16 of Neptune. These satellites are further categorized into regular and irregular groups based on the Laplace radius of their respective host planets \citep{Tremaine2009Satellite}:
\begin{equation}
r_{\rm L} = \left(J_2 R_{\rm p}^2 a_{\rm p}^3 (1-e_{\rm p}^2)^{3/2} \frac{m_{\rm p}}{m_{\odot}}\right)^{1/5},
\end{equation}
where $J_2$ is the quadrupole distortion coefficient, and $R_{\rm p}$, $a_{\rm p}$, $e_{\rm p}$ and $m_{\rm p}$ are the host planet's radius, semimajor axis, eccentricity and mass, and ${m_{\odot}}$ stands for the solar mass. The values of $J_2$, which account for the contribution from the inner satellite system, are adopted from \citet{Tremaine2009Satellite}. At the Laplace radius $r_{\rm L}$, a dynamic balance is achieved between the perturbations from the host planet’s non-sphericity and the solar third-body gravitational force. Specifically, in the inner region ($r<r_{\rm L}$), the dynamics is dominated by the quadrupole potential of the host planet’s oblateness, whereas in the outer region ($r>r_{\rm L}$), the solar quadrupole perturbation becomes the dominant influence \citep{grishin2024irregularII}.

\begin{figure}
\centering
\includegraphics[width=0.95\columnwidth]{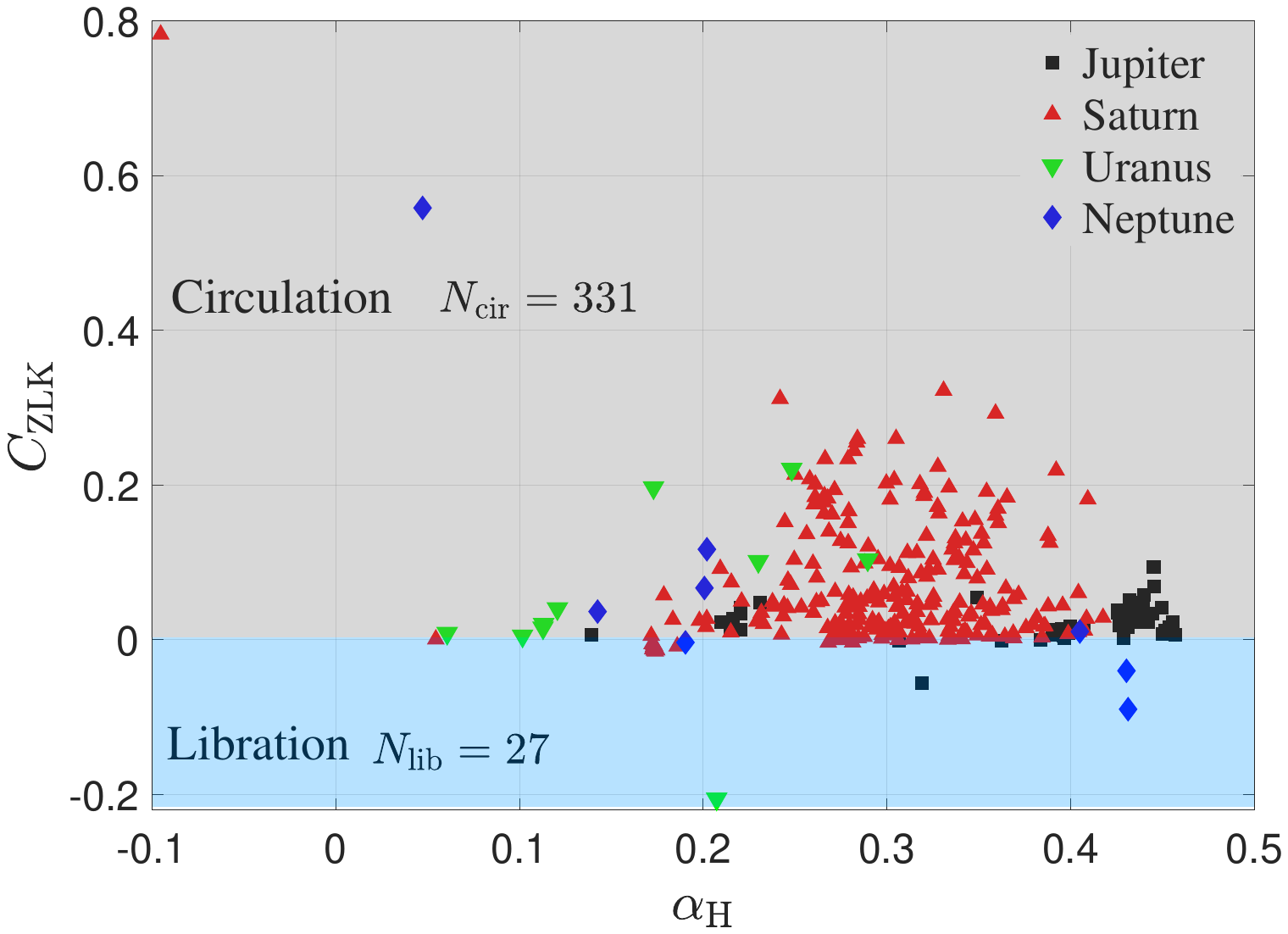}
\includegraphics[width=0.95\columnwidth]{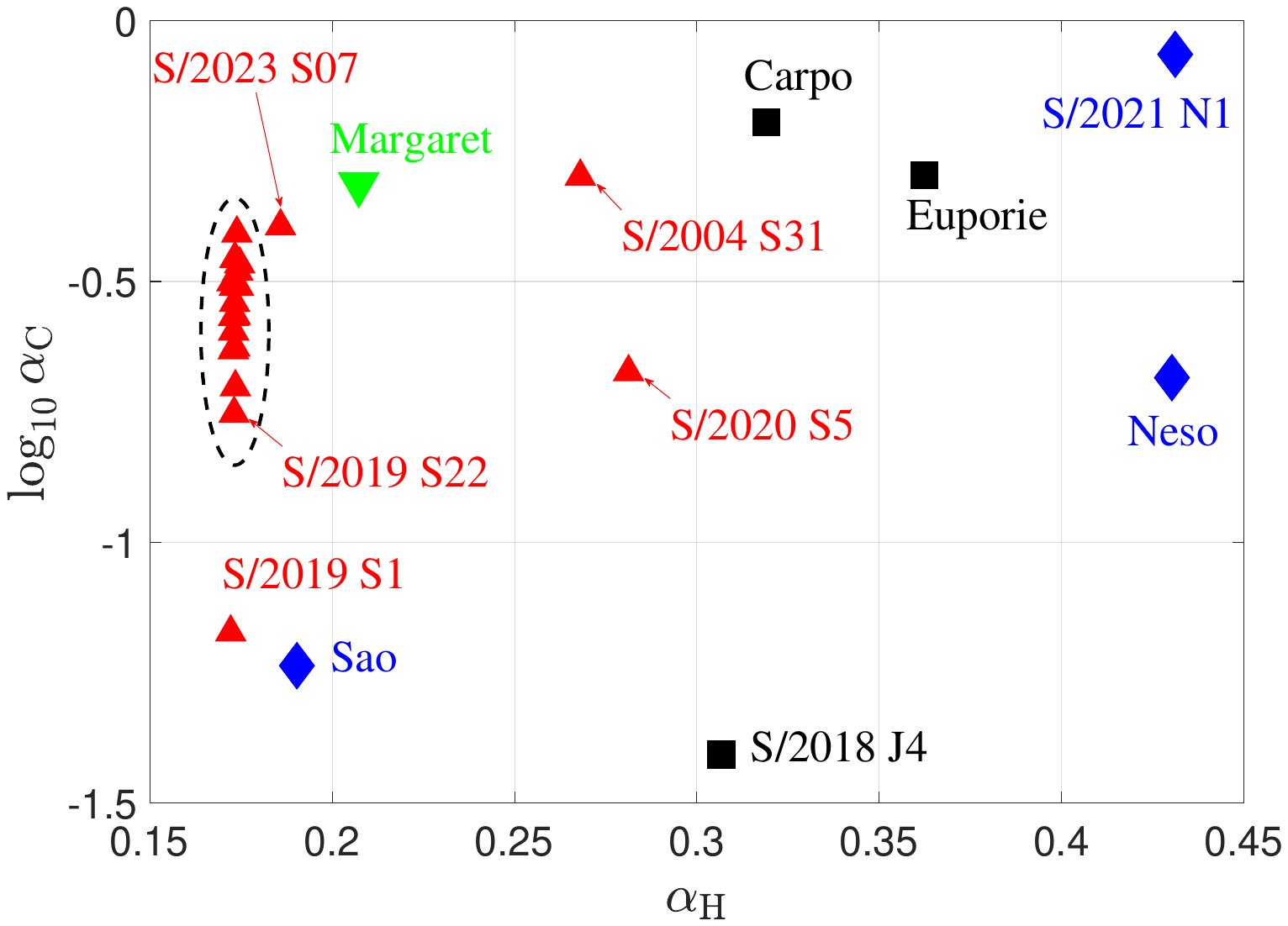}
\caption{Distribution of 358 irregular satellites of giant planets in the space of $(\alpha_{\rm H},C_{\rm ZLK})$ (\textit{top panel}), as well as the associated distribution of 27 candidates of librating satellites (\textit{bottom panel}). In the top panel, the irregular satellites are divided into two groups by the separatrix of $C_{\rm ZLK}=0$: librating satellites located in the region of $C_{\rm ZLK}<0$ (27 satellites) and circulating satellites located in the region of $C_{\rm ZLK}>0$ (331 satellites). The bottom panel displays a group of librating Saturnian satellites, enclosed by the dashed curve, with S/2019 S22 as a representative member. In both panels, the Jovian, Saturnian, Uranian, and Neptunian satellites are indicated in black, red, green, and blue, respectively.}
\label{Fig3}
\end{figure}

According to the Laplace radius $r_{\rm L}$, satellites are categorized as irregular if their semi-major axis $a$ exceeds $r_{\rm L}$, and regular if $a < r_{\rm L}$. Applying this criterion, 358 irregular satellites (ISs) are identified among the 413 known satellites of giant planets. Specifically, the population of ISs includes 89 around Jupiter, 251 around Saturn, 10 around Uranus, and 8 around Neptune. Different definitions used to distinguish irregular satellites from regular ones can be found in \citet{burns1986some}. However, different definitions give the same statistical result because of the sharp distinction between regular and irregular satellites \citep{jewitt2007irregular}.

It should be noted that our extended Brown Hamiltonian is formulated based on mean orbital elements. Therefore, applying the previously discussed criterion requires the derivation of the corresponding mean orbital elements for the satellites. To this end, a numerical-averaging approach is adopted \citep{grishin2024irregularII}. First, the initial states of satellites at the epoch of October 18, 2024, are retrieved from the NASA Horizons system \footnote{Please see \url{https://ssd.jpl.nasa.gov/horizons/}.}. Second, these states are transformed into the Sun–Planet invariant plane reference frame to maintain consistency with the Brown Hamiltonian framework. Third, we perform $N$-body simulations and average the orbital elements over one orbital period of the outer binary. For brevity, the time-averaged orbital elements of satellites are hereafter denoted as $(a,e,i,\Omega,\omega)$. As a result, it becomes possible to determine the motion integral of each satellite, including $\alpha_{\rm H}$, $j_z$ (or the equivalent parameter $i_*$) and $C_{\rm ZLK}$. These conserved parameters are important to characterize their long-term dynamical structures.

Using the database of time-averaged elements, we map the distribution of the 358 irregular satellites in the parameter space $(\alpha_{\rm H}, C_{\rm ZLK})$, as shown in the top panel of Figure \ref{Fig3}. It shows that 331 satellites have $C_{\rm ZLK} > 0$, whereas 27 satellites satisfy $C_{\rm ZLK} < 0$. Based on the analytical criterion established in the preceding section, these 27 satellites with $C_{\rm ZLK} < 0$ are identified as candidates currently trapped in the ZLK resonance. In particular, their distribution is as follows: 3 orbiting Jupiter, 20 orbiting Saturn, 1 orbiting Uranus, and 3 orbiting Neptune.

To facilitate our analysis, we introduce the normalized integral,
\begin{equation}
\alpha_{\rm C} = C_{\rm ZLK} / C_{\rm ZLK}^{\rm fix},
\end{equation}
where $C_{\rm ZLK}^{\rm fix}$ is the Lidov integral evaluated at the modified ZLK center. By definition, $\alpha_{\rm C}$ ranges from 0 (at the separatrix) to 1 (at the ZLK center). This parameter effectively measures the depth of resonance: a higher $\alpha_{\rm C}$ corresponds to a more deeply trapped state within the ZLK resonance, whereas a lower $\alpha_{\rm C}$ marks a trajectory closer to the %edge of the libration regime.
separatrix. The distribution of these 27 candidate satellites is further illustrated in the $(\alpha_{\rm H}, \log_{10} \alpha_{\rm C})$ plane, as shown in Figure \ref{Fig3} (bottom). We see that numerous librating satellites of Saturn (indicated by red dots) occupy remarkably similar orbits. This clustering suggests they are likely fragments resulting from the collisional disruption of larger, captured progenitors \citep{peale1999origin,gladman2001discovery}.

\begin{figure*}
\raggedright
\includegraphics[width=0.66\columnwidth]{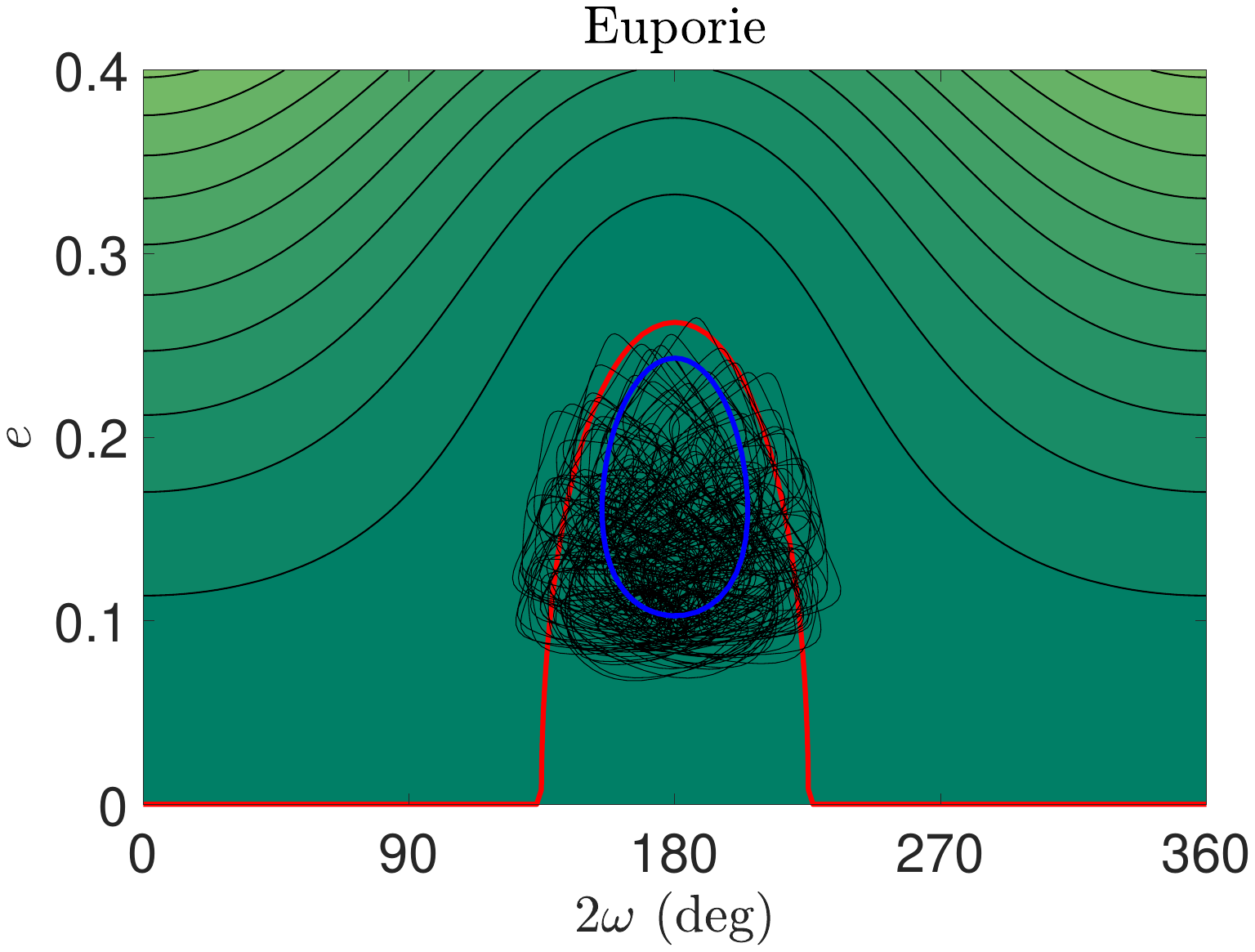}
\includegraphics[width=0.66\columnwidth]{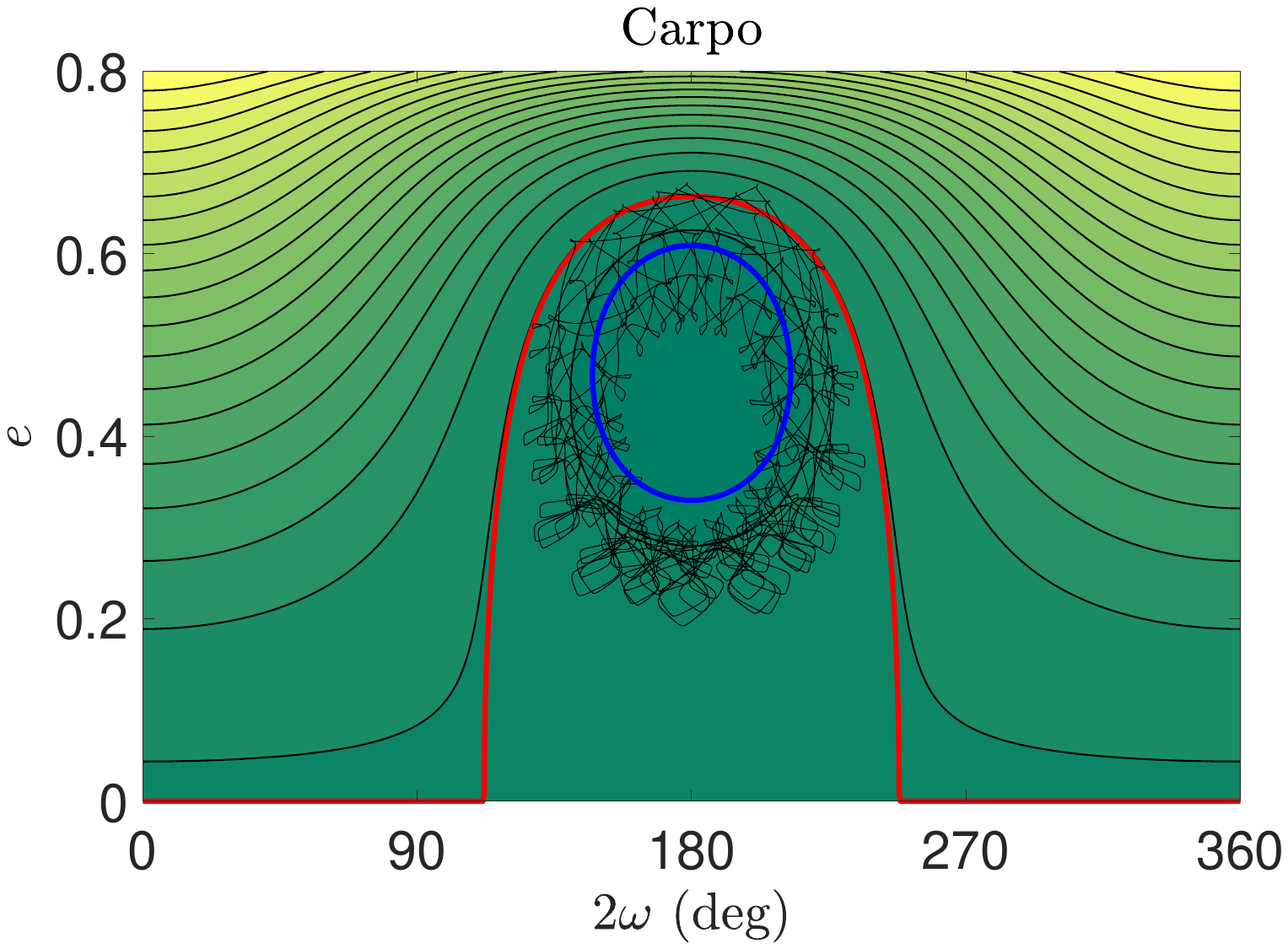}
\includegraphics[width=0.66\columnwidth]{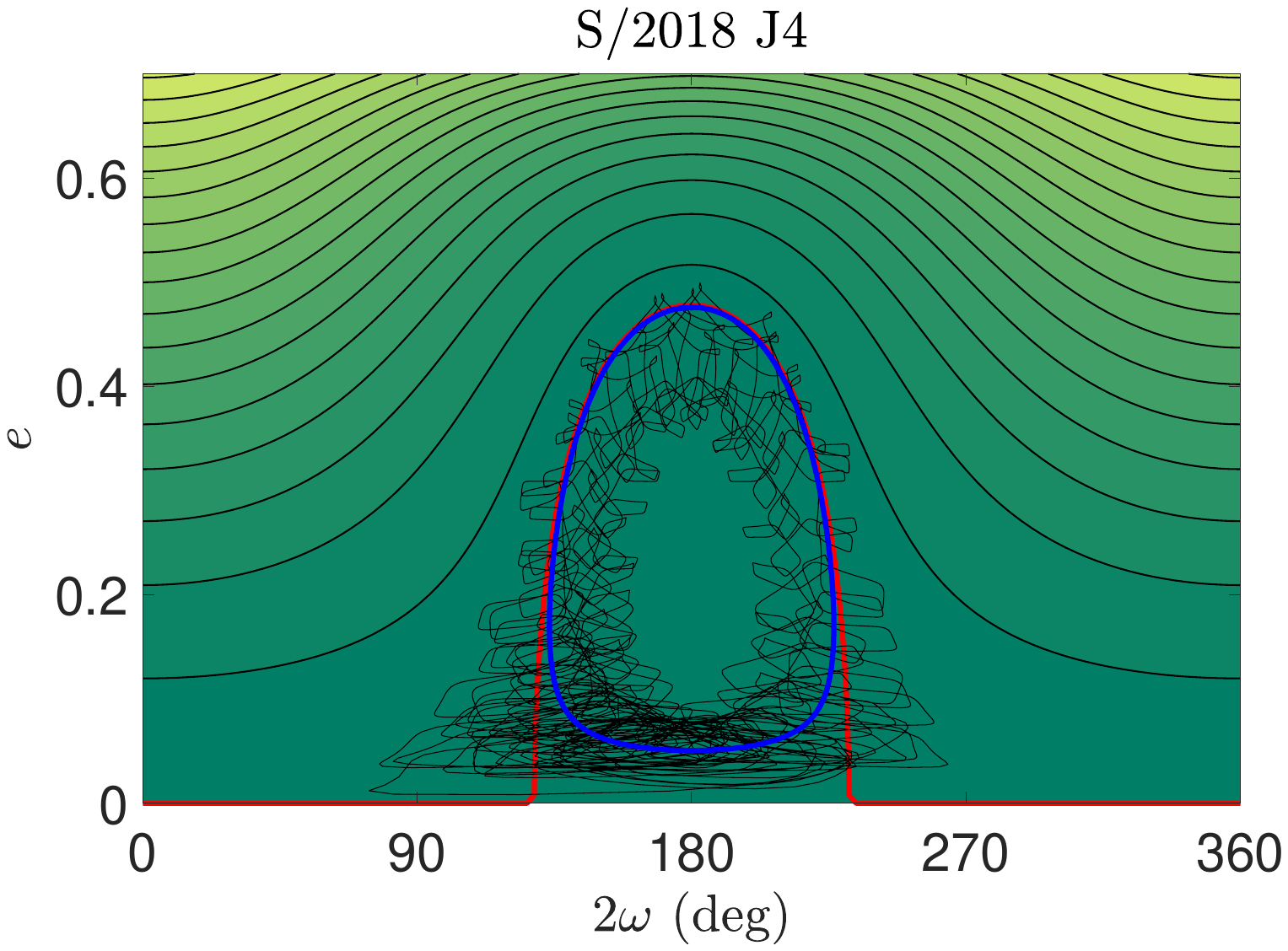}\\
\includegraphics[width=0.66\columnwidth]{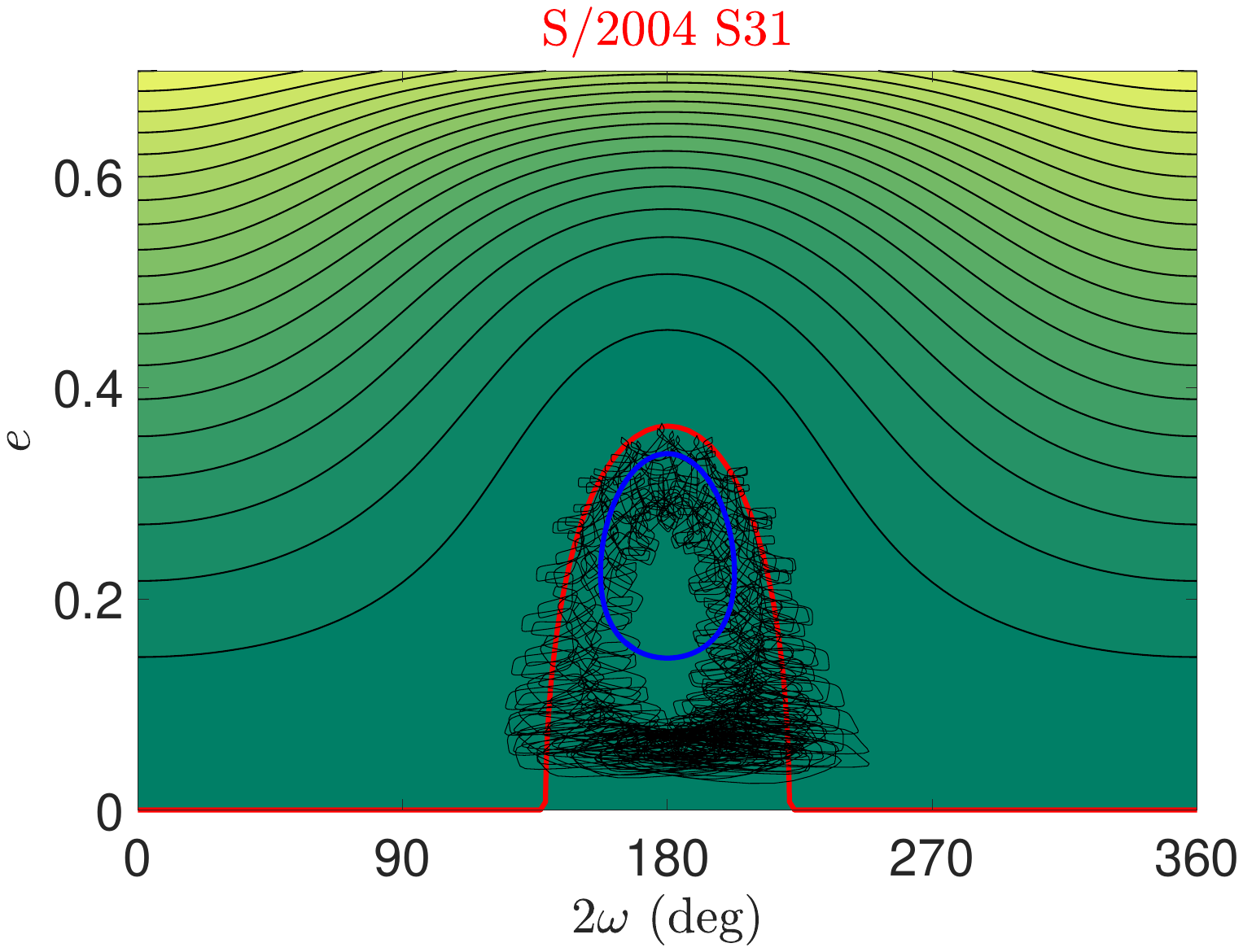}
\includegraphics[width=0.66\columnwidth]{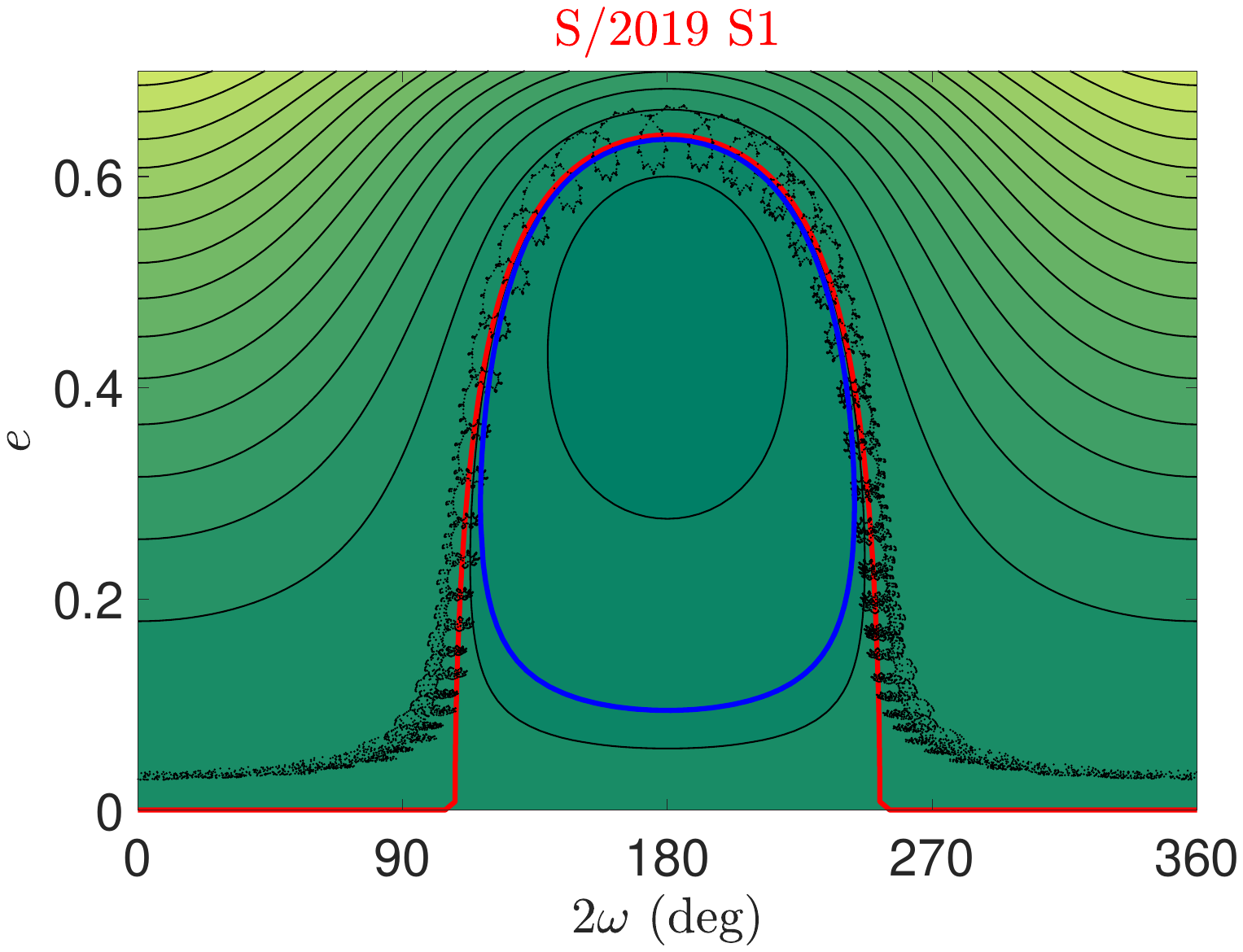}
\includegraphics[width=0.66\columnwidth]{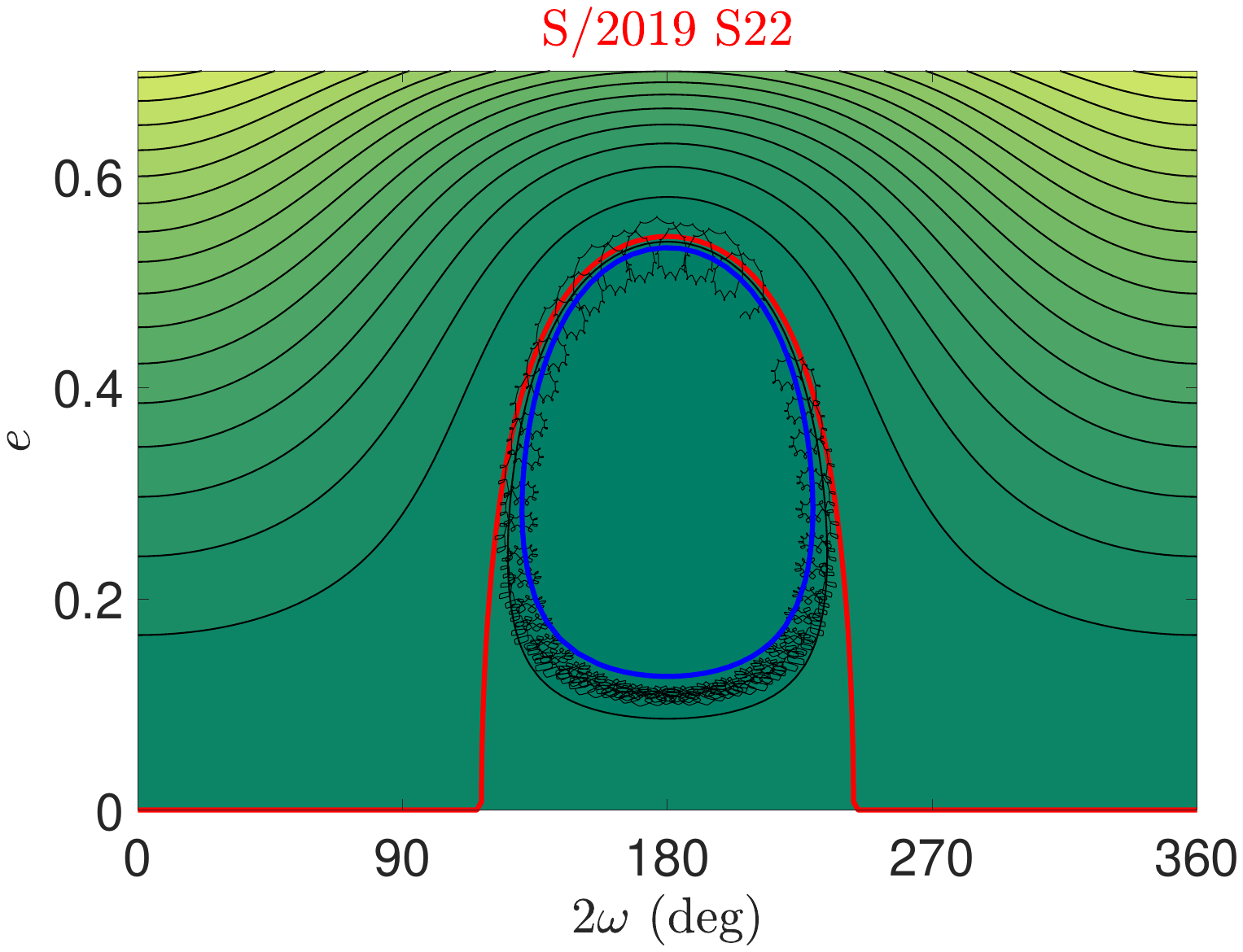}\\
\includegraphics[width=0.66\columnwidth]{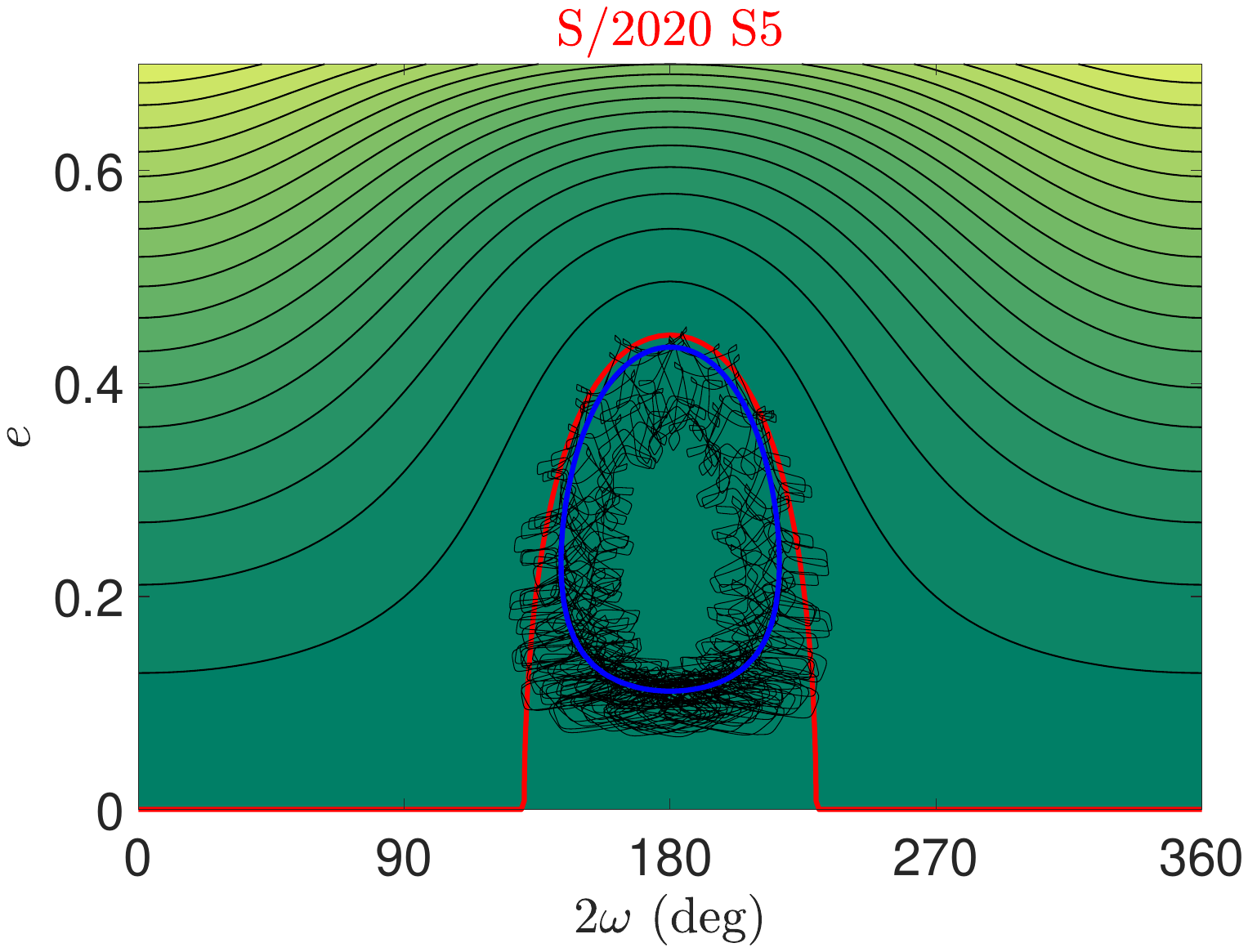}
\includegraphics[width=0.66\columnwidth]{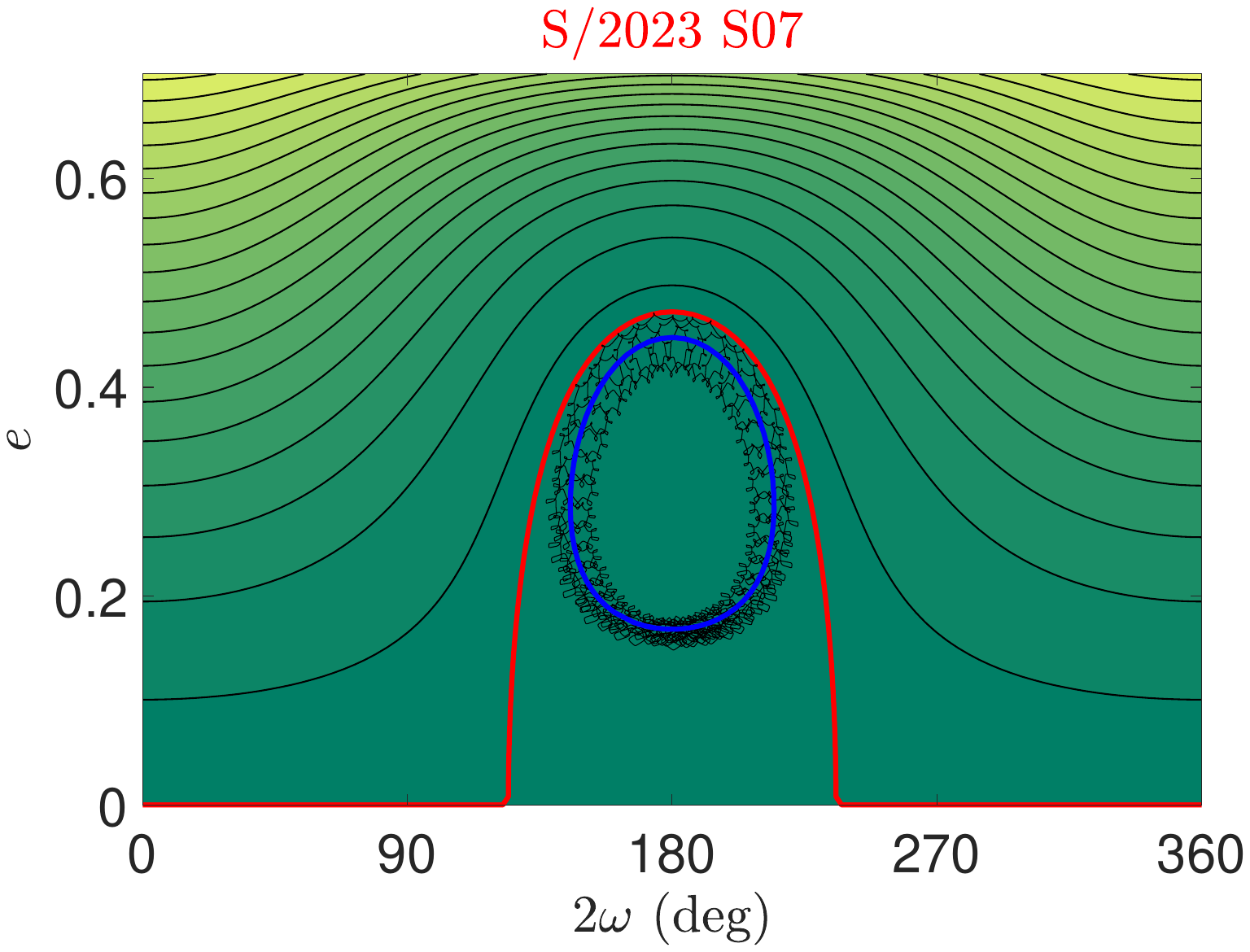}
\includegraphics[width=0.66\columnwidth]{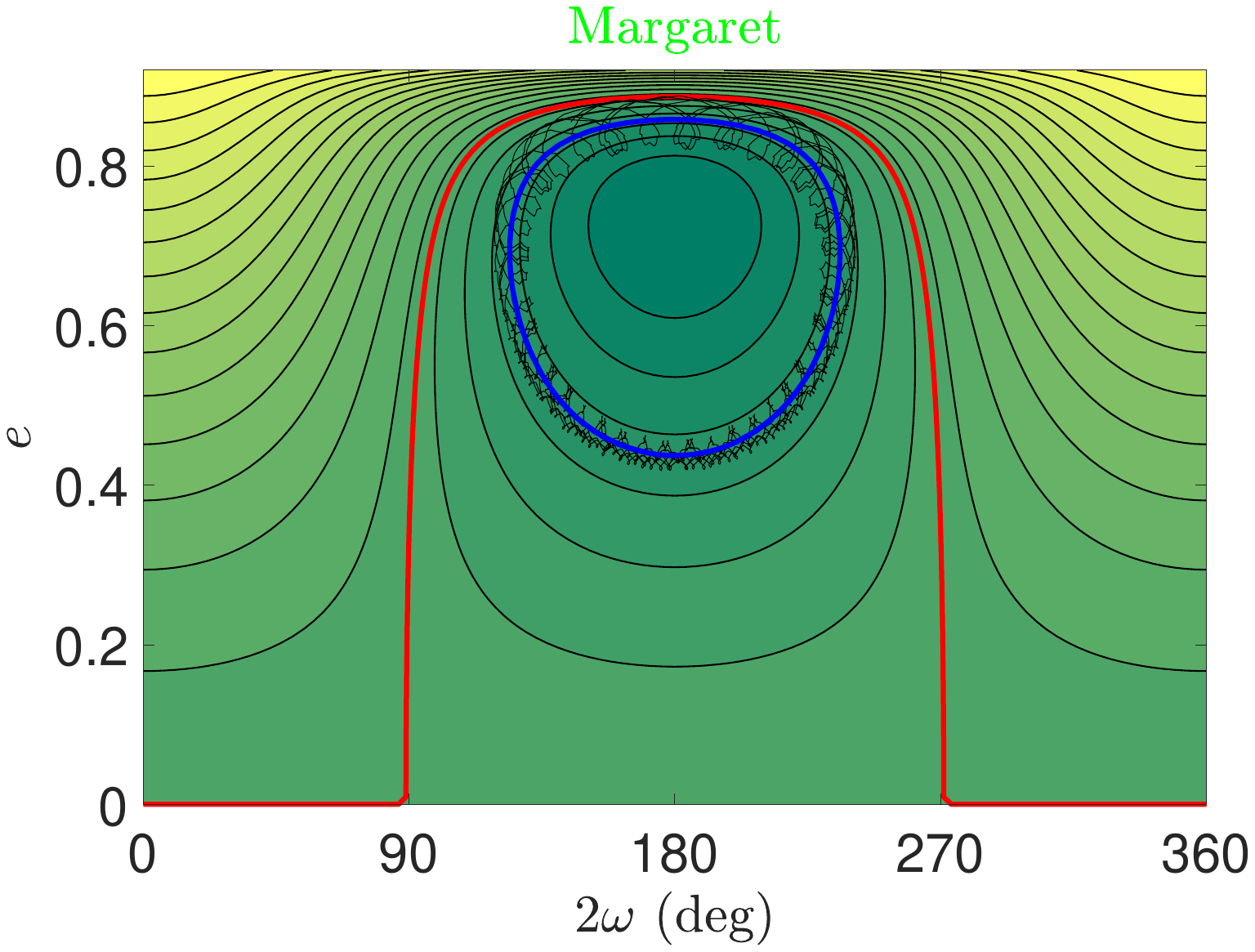}\\
\includegraphics[width=0.66\columnwidth]{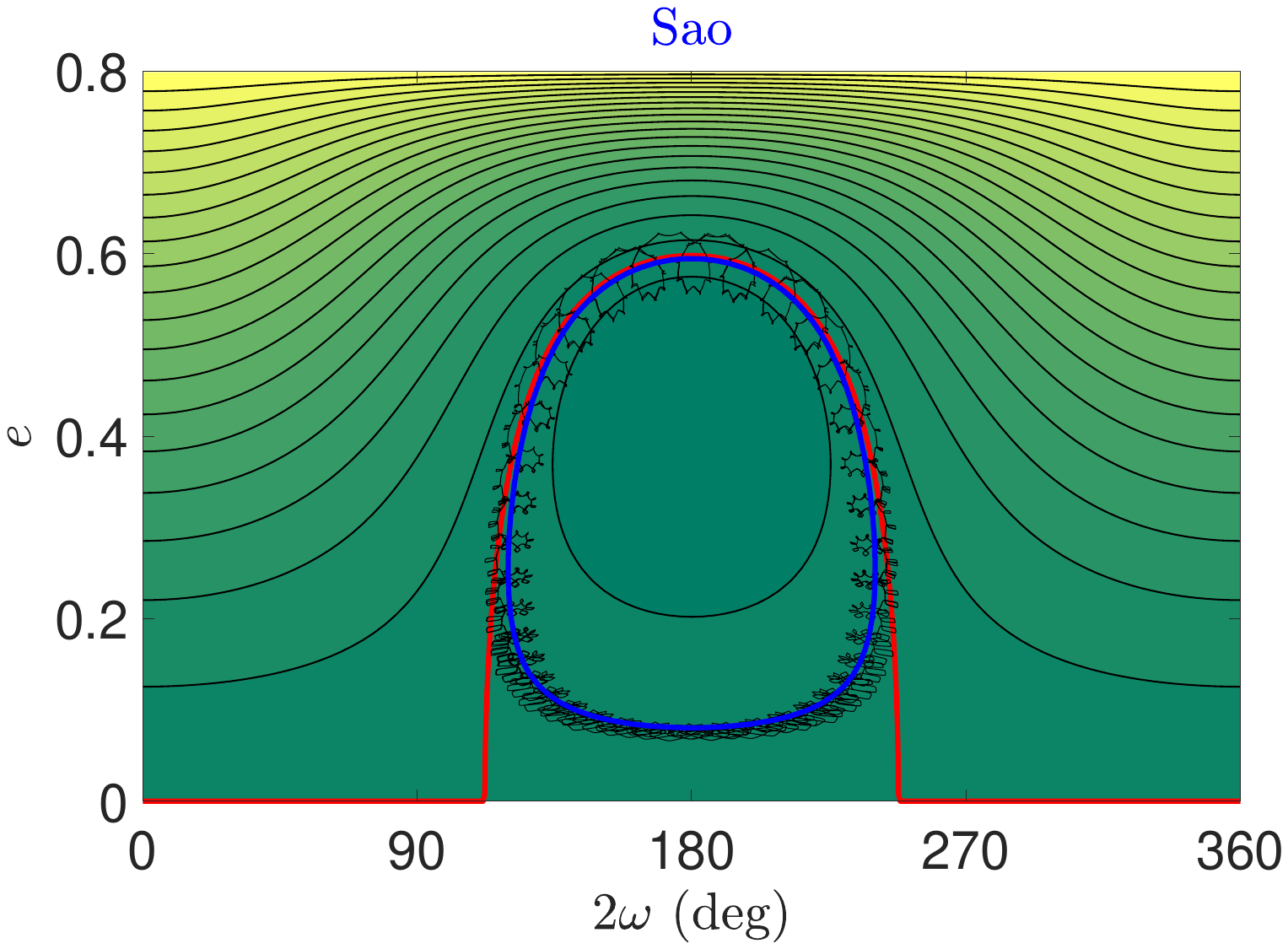}
\includegraphics[width=0.66\columnwidth]{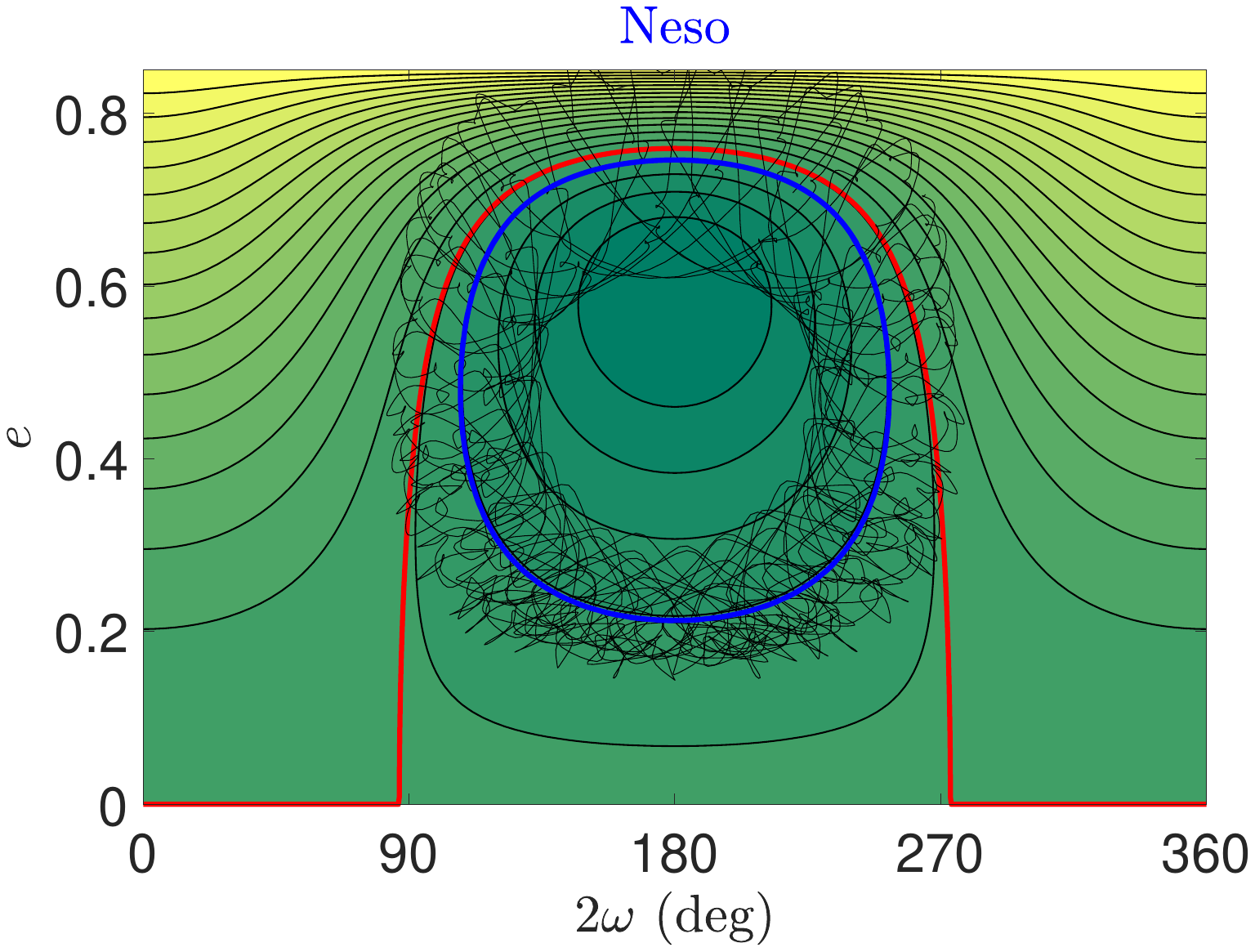}
\includegraphics[width=0.66\columnwidth]{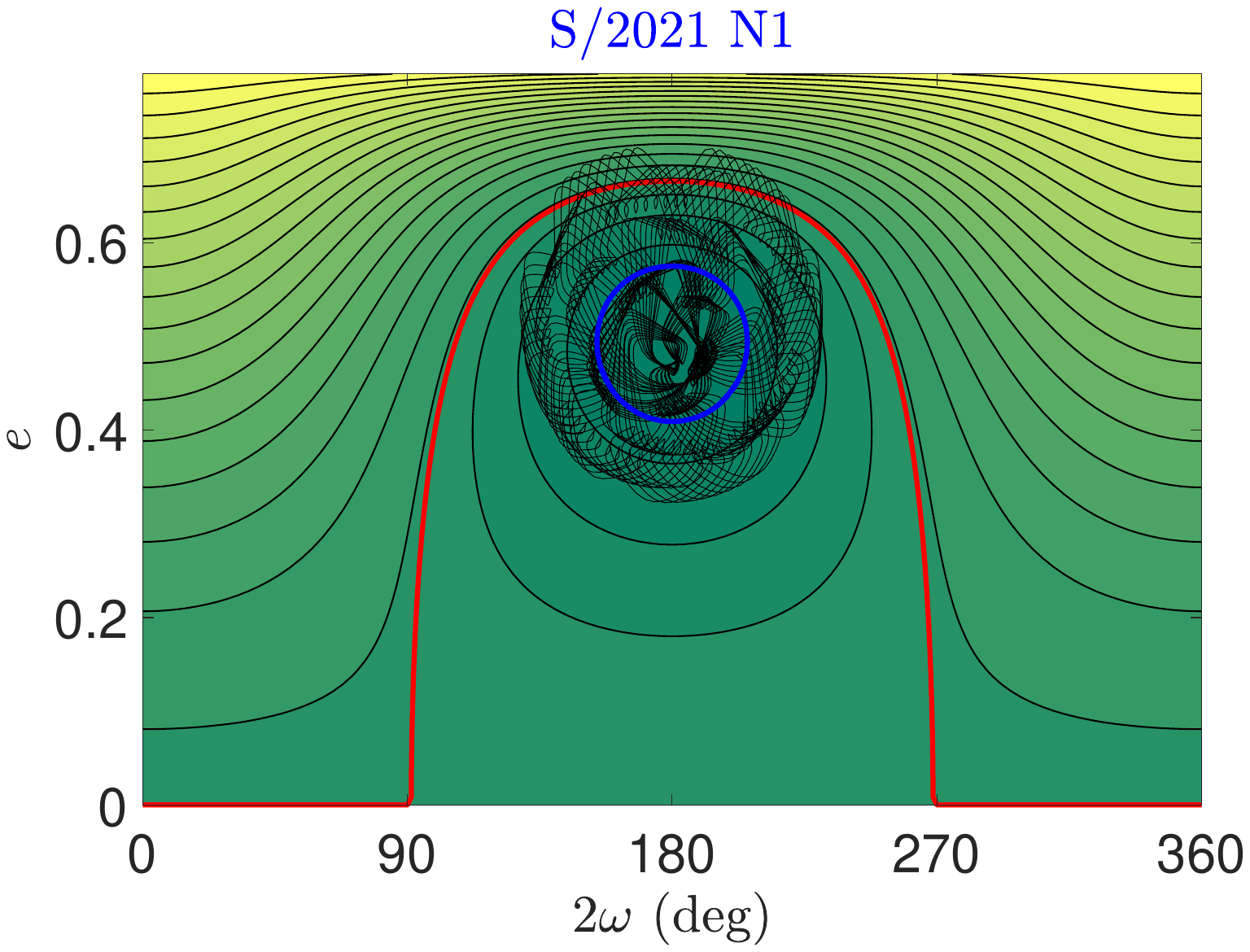}
\caption{Osculating trajectories propagated under the full three-body model (black lines), averaged orbits propagated under the extended Brown Hamiltonian model (blue lines), together with the associated phase portraits (level curves of $C_{\rm ZLK}$) as the background, for those candidate librating satellites shown in the bottom panel of Figure \ref{Fig3}. Numerical simulations of the remaining satellites [S/2004 (S54, S55), S/2005 S4, S/2007 S10, S/2019 (S23, S24, S25, S26), S/2020 (S1, S11, S12, S13), S/2023 S02, Ijiraq, and Kiviuq] confirm they share similar dynamical behaviours with S/2019 S22 (they belong to the group highlighted in the bottom panel of Fig. \ref{Fig3}). For brevity, these results are not shown here. Almost all satellites exhibit a good consistency between their osculating and averaged orbits.}
\label{Fig4}
\end{figure*}

To validate the analytical results presented in Figure \ref{Fig3}, we conduct direct three-body integrations for all 27 candidate satellites identified within the ZLK resonances \footnote{The introduction of additional giant planets to the dynamical model shows no difference within the time-scale considered in the present paper. However, for much longer time-scales it may introduce deviation \citep{saha1993orbits,grishin2024irregularII}.}. Their osculating trajectories are plotted as black lines in Figure \ref{Fig4}. Simultaneously, their averaged orbits are produced under the extended Brown Hamiltonian model (see the blue lines). In each panel, these trajectories are overlaid on their associated phase portraits (i.e., level curves of $C_{\rm ZLK}$), where the dynamical separatrix is highlighted by red lines.

In general, the result shows good agreement between the osculating and averaged trajectories for each satellite, confirming that the extended Brown Hamiltonian is sufficiently robust to predict the long-term evolution of irregular satellites. In particular, we can see that, except for S/2019 S1, all the remaining candidates are verified to be trapped within the ZLK resonance. The discrepancy observed for S/2019 S1 is attributed to its proximity to the dynamical separatrix, a region characterized by chaotic dynamics resulting from resonant interaction \citep{carruba2004chaos}. Accordingly, we confirm 26 satellites in total to reside in the ZLK resonance. Notice that half of these objects have been reported by \citet{grishin2024irregularII}, who utilized a smaller data set in their analysis.

In Jupiter's system, it is verified that there are 3 irregular satellites inside the ZLK resonance: Euporie, Carpo and S/2018 J4. The ZLK libration of Euporie and Carpo was reported by \citet{brozovic2017orbits}, and the libration of S/2018 J4 was stated by \citet{grishin2024irregularII}. The bottom panel of Figure \ref{Fig3} {shows that} these  librating satellites occupy distant orbits around Jupiter with $\alpha_{\rm H}>0.3$, exhibiting large-scale oscillations during their long-term evolutions, as illustrated in the first three panels of Figure \ref{Fig4}. Regarding secular models, \citet{grishin2024irregularII} has realized that the Brown Hamiltonian model developed in \citet{luo2016double} is not accurate enough to predict the long-term behaviors of Euporie and Carpo, because they have relatively high value of $\varepsilon_{\rm SA}$ (single-averaging parameter), showing a weak-level hierarchy of system. In particular, their Figure 4 shows a significant discrepancy between its osculating trajectory under the full three-body model and the averaged trajectory under the classical Brown Hamiltonian model (corresponding to ${\cal F}_{20}+{\cal F}_{21}$). For this type of weak-hierarchy systems, inclusion of the extended Brown correction (${\cal F}_{22}$) is necessary and it can greatly improve the accuracy of long-term prediction (see examples in Paper I). In addition, although it is verified that S/2018 J4 lies in the ZLK resonance, it is close to the separatrix. This observation is consistent with that in \citet{grishin2024irregularII}. Thus, its motion is nearly chaotic, and it may escape from the current resonant state in the future with the consideration of ignored perturbations (e.g., perturbations from other planets).

In Saturn's system, there are 19 irregular satellites trapped inside the ZLK resonance. According to their distribution in the space of $(\alpha_{\rm H},\alpha_{\rm C})$, there are two groups for Saturn's librating satellites. The members of the first group, occupying distant orbits with $\alpha_{\rm H}$ greater than 0.25, present large-amplitude periodic oscillations, including S/2004 S31 and S/2020 S5. The ones of the second group, occupying relatively close orbits with $\alpha_{\rm H}$ smaller than 0.2, exhibit regular long-term behaviors with small periodic oscillations and share similar phase-space structures, including S/2004 S54, S/2004 S55, S/2005 S4, S/2007 S10, S/2019 S22, S/2019 S23, S/2019 S24, S/2019 S25, S/2019 S26, S/2020 S1, S/2020 S11, S/2020 S12, Ijiraq, Kiviuq and S/2023 S7.

Margaret is the only one librating satellite in Uranus's system \citep{brozovic2009orbits,brozovic2022orbits,grishin2024irregularII}. It is moving on a prograde and high-eccentricity orbit. For this satellite, the usual secular model without any Brown corrections can work well in predicting its long-term evolution \citep{grishin2024irregularII}. This is because of its low $j_z\approx 0.348$ and small $\alpha_{\rm H}\approx 0.2$, which make the contribution from ${\cal F}_{21}$ and ${\cal F}_{22}$ negligible. During the ZLK cycles, the minimum eccentricity is greater than 0.4 and its maximum eccentricity can reach up to 0.9. However, its orbit is highly stable because it is trapped in deep ZLK resonance. For this satellite, the Hamiltonian perturbation theory based on the elliptic expansion of disturbing function may fail to work due to the Laplace limit of $e_{\rm c} = 0.6627$ \citep{wintner1941analytical}.

Neptune hosts three librating satellites, including Sao, Neso and S/2021 N1. Notice that the libration of Sao and Neso was reported in \citet{brozovic2011orbits} and \citet{brozovic2022orbits}, and the libration of S/2021 N1 was presented in \citet{grishin2024irregularII}. According to the bottom panel of Figure \ref{Fig3}, we can see that the satellite Sao has a small separation $\alpha_{\rm H}$, while Neso and S/2021 N1 move on distant orbits with $\alpha_{\rm H}$ greater than 0.4. Thus, Sao has a relatively regular evolution, while Neso and S/2021 N1 exhibit significant short-term oscillations during the long-term evolution. In particular, the eccentricity of Neso can be excited up to 0.8. The librating satellite S/2021 N1 is trapped deeply inside the ZLK resonance because its $\alpha_{\rm C}$ is close to unity, as shown in the bottom panel of Figure \ref{Fig3}.   

\section{Conclusions}
\label{Sect5} 

We applied the extended Brown Hamiltonian and the modified Lidov integral to the irregular satellites of the giant planets in our Solar System. Under this framework, the long-term orbital evolution is governed by three conserved quantities: the semimajor axis in Hill radii units $\alpha_{\rm H}$, the vertical angular momentum $j_z$, and the Hamiltonian constant ${\cal F}$ (or the Lidov integral $C_{\rm ZLK}$ and $\alpha_{\rm C}$). These constants of motion are expressed in terms of time-averaged orbital elements, and %they provide fundamental parameters for
govern the long-term dynamical analysis: The separation $\alpha_{\rm H}$ is related to the single-averaging parameter $\varepsilon_{\rm SA}$ (or $\varepsilon_{21}$), measuring the level of timescale hierarchy; $j_z$ determines the eccentricity of the ZLK center; and the parameter $\alpha_{\rm C}$ measures the depth of the ZLK resonance.  

The motion modes of satellites in the phase space are divided by the dynamical separatrix of $C_{\rm ZLK}=0$. Specifically, $C_{\rm ZLK}>0$ characterizes the ZLK circulation regime, while $C_{\rm ZLK}<0$ identifies the ZLK libration. Applying this analytical criterion to 358 irregular satellites, we categorize 331 into the circulation regime and 27 into the libration regime. Direct $N$-body simulations validate these predictions, confirming that 26 of the 27 candidates reside within the ZLK resonance. The sole exception is the Saturnian satellite S/2019 S1, because it is located in close proximity to the separatrix.

To conclude, the criterion $C_{\rm ZLK}<0$ provides an efficient and robust method for identifying candidates trapped within the ZLK resonance. As a fundamental parameter, the modified Lidov integral $C_{\rm ZLK}$ is crucial for characterizing the ZLK dynamics of irregular satellites. Importantly, the applicability of this analysis extends beyond solar system dynamics, offering a similarly effective approach for identifying orbital families of binary black holes in the Galactic Center \citep{grishin26bbh}.

\section*{Acknowledgements}
We are grateful to Prof. Valerio Carruba for his expert review and insightful suggestions, which have substantially improved this work. Hanlun Lei and Xiaoyan Leng are financially supported by the National Natural Science Foundation of China (Nos. 12573063 and 12233003) and the China Manned Space Program with grant no. CMS-CSST-2025-A16. Evgeni Grishin acknowledges support from the ARC Discovery Program DP240103174 (PI: Heger), the ARC OzGrav Centre of Excellence CE230100016, and the ARC Discovery Early Career Research Award (DECRA) DE260101802.
%%%%%%%%%%%%%%%%%%%%%%%%%%%%%%%%%%%%%%%%%%%%%%%%%%

\section*{Data Availability}
The codes used in this article could be shared on reasonable request. The original (osculating) orbit elements of Jupiter's irregular satellites are extracted from Horizons system (https://ssd.jpl.nasa.gov/horizons/).
\bibliographystyle{mnras}
\bibliography{mybib} % if your bibtex file is called example.bib
%%%%%%%%%%%%%%%%%%%%%%%%%%%%%%%%%%%%%%%%%%%%%%%%%%
%%%%%%%%%%%%%%%%% APPENDICES %%%%%%%%%%%%%%%%%%%%%
% \appendix
% \section{Some extra material}
% If you want to present additional material which would interrupt the flow of the main paper, it can be placed in an Appendix which appears after the list of references.
%%%%%%%%%%%%%%%%%%%%%%%%%%%%%%%%%%%%%%%%%%%%%%%%%%
% Don't change these lines
\bsp	% typesetting comment
\label{lastpage}
\end{document}